\documentclass[aps,nofootinbib,prd,superscriptaddress,tightenlines,notitlepage,twocolumn,showpacs]{revtex4}

\usepackage[colorlinks]{hyperref}
\usepackage{tabularx}
\usepackage{graphicx}
\usepackage{amssymb,bm}
\usepackage{xcolor}
\usepackage{amsmath}
\usepackage[varg]{txfonts}
\usepackage{enumerate}
\usepackage[normalem]{ulem}
\usepackage{subfigure}
\usepackage{dcolumn}
\usepackage{bm}
\usepackage{multirow}
\newcommand{\apjl}{Astrophys. J. Lett.}
\newcommand{\apjs}{Astrophys. J. Suppl. Ser.}
\newcommand{\aap}{Astron. \& Astrophys.}
\newcommand{\aj}{Astron. J.}
\newcommand{\mnras}{Mon. Not. R. Astron. Soc.}
\newcommand{\jcap}{JCAP}

\newcommand{\pasp}{Pub. Astro. Soc. Pacific}

\newcommand{\PMO}{\affiliation{Purple Mountain Observatory, Chinese Academy of Sciences, Nanjing 210023, China}}
\newcommand{\USTC}{\affiliation{School of Astronomy and Space Sciences, University of Science and Technology of China, Hefei 230026, China}}

\begin{document}


\title{Cosmology-independent Photon Mass Limits from Localized Fast Radio Bursts by using Artificial Neural Networks}

\author{Jing-Yu Ran}\thanks{These authors contributed equally.}
\PMO\USTC

\author{Bao Wang}\thanks{These authors contributed equally.}
\PMO\USTC

\author{Jun-Jie Wei}\thanks{jjwei@pmo.ac.cn}\PMO\USTC





\date{\today}

\begin{abstract}
A hypothetical photon mass, $m_{\gamma}$, can produce a frequency-dependent vacuum dispersion of light, which leads to an additional time delay between photons with different frequencies when they propagate through a fixed distance. The dispersion measure--redshift measurements of fast radio bursts (FRBs) have been widely used to constrain the rest mass of the photon. However, all current studies analyzed the effect of the frequency-dependent dispersion for massive photons in the standard $\Lambda$CDM cosmological context. In order to alleviate the circularity problem induced by the presumption of a specific cosmological model based on the fundamental postulate of the masslessness of photons, here we employ a new model-independent smoothing technique, Artificial Neural Network (ANN), to reconstruct the Hubble parameter $H(z)$ function from 34 cosmic-chronometer measurements. By combining observations of 32 well-localized FRBs and the $H(z)$ function reconstructed by ANN,
we obtain an upper limit of $m_{\gamma} \le 3.5 \times 10^{-51}\;\rm{kg}$, or equivalently $m_{\gamma} \le 2.0 \times 10^{-15}\;\rm{eV/c^2}$ ($m_{\gamma} \le 6.5 \times 10^{-51}\;\rm{kg}$, or equivalently $m_{\gamma} \le 3.6 \times 10^{-15}\;\rm{eV/c^2}$) at the $1\sigma$ ($2\sigma$) confidence level. This is the first cosmology-independent photon mass limit derived from extragalactic sources.
\end{abstract}

\maketitle


\section{\label{sec:introduction}Introduction}

As a fundamental postulate of Maxwell's electromagnetism and Einstein's special relativity, the principle of the constancy of the speed of light implies the masslessness of photons, as elucidated by the particle-wave duality. This principle is also embraced within the framework of general relativity (GR). However, even a very minute photon mass, if existent, would necessitate new physical theories, such as the renowned de Broglie-Proca theory \citep{De1922, Proca1936}, the model of massive photons as an explanation for dark energy \citep{2016PhRvD..93h3012K}, and other new ideas in the standard-model extension with massive photons \citep{2021EPJC...81....4S}. Consequently, achieving precise constraints on the rest mass of photon, denoted as $m_{\gamma}$, remains an imperative.

Numerous experimental and observational constraints on $m_{\gamma}$ have been derived from various effects resulting from the hypothetical nonzero photon mass. These include tests of Coulomb's inverse square law \citep{Williams1971}, the Cavendish torsion balance \citep{Lakes1998,Luo2003}, gravitational deflection of electromagnetic waves \citep{Lowenthal1973}, mechanical stability of magnetized gas in galaxies \citep{Chibisov1976}, magneto-hydrodynamic phenomena of the solar wind \citep{Ryutov1997,Ryutov2007,2012ChPhL..29k1401L,Retin2016}, Jupiter’s magnetic field \citep{Davis1975}, the spindown of a pulsar \citep{Yang2017}, among others. However, these constraints are contingent upon specific theories of massive photons and are thus dynamic tests in nature.

In contrast, a kinematic test based on the dispersion effect of the speed of light in vacuum is more inherently pure \citep{2005ChPhL..22.3057T,2005RPPh...68...77T}. The dispersion relation for massive photons is governed by
\begin{equation}
E^2 = p^2c^2 + m_{\gamma}^2c^4\;,
\end{equation}
where $p$ represents the momentum. The group velocity of a photon with energy $E=h\nu$ can be expressed as
\begin{equation}\label{2}
v = \frac{\partial E}{\partial p} = c\sqrt{1-\frac{m_{\gamma}^2c^4}{E^2}} \approx c\left ( 1-\frac{1}{2}\frac{m_{\gamma}^2c^4}{h^2\nu ^2} \right )\;,
\end{equation}
where $h$ is the Planck constant and the last term is valid when $m_{\gamma}\ll h\nu/c^2\simeq 7\times 10^{-42}\mathrm{(\nu /GHz)\;kg}$. Eq.~(\ref{2}) suggests that if two photons with different frequencies are emitted simultaneously from the same source, they will be observed at different times due to their different velocities.
Fast radio bursts (FRBs) provide an excellent celestial laboratory for detecting this dispersion effect, owing to their characteristics of (i) short time durations, (ii) long propagation distances, and (iii) low-frequency emissions.

Based on the vacuum dispersion method, several studies have used FRBs to put stringent upper limits on the photon mass
\citep{2016ApJ...822L..15W,2016PhLB..757..548B,2017PhLB..768..326B,2017PhRvD..95l3010S,2019ApJ...882L..13X,
2020RAA....20..206W,2021PhLB..82036596W,2023JCAP...01..010C,Lin2023,2023JCAP...09..025W,2024ApJ...965...38W}.
Since FRBs originate at cosmological distances,
the cosmic expansion rate $H(z)$ has to be considered in constraining the photon mass $m_{\gamma}$.
In all previous studies, the required $H(z)$ information is calculated within the standard $\Lambda$CDM cosmological model. It should, however, be emphasized that $\Lambda$CDM is rooted in the framework of GR, which also embraces the postulate of the constancy of light speed. Thus, there is a circularity problem in constraining the photon mass.
To address this problem, one has to determine the Hubble parameter $H(z)$ in a cosmology-independent way.

In this work, we propose a novel nonparametric approach, utilizing the Artificial Neural Network (ANN) technology, to
reconstruct a smooth $H(z)$ function that best approximates the discrete $H(z)$ measurements.
By combining observations of well-localized FRBs and the $H(z)$ function reconstructed by ANN,
we aim to establish a cosmology-independent constraint on the rest mass of the photon.
The paper is structured as follows. Section \ref{sec:method} introduces the theoretical framework and the observational data utilized for constraining the photon mass. Section \ref{sec:results} presents the cosmology-independent constraints on the photon mass and other relevant parameters. Finally, a brief summary and discussions are given in Section \ref{sec:discussions}.

\section{\label{sec:method}Analysis Method and Data}
\subsection{\label{sec:plasma}Theoretical framework}
In our analysis, the observed time delay between different frequencies for FRBs is attributed to two main factors: (i) the propagation of photons through the plasma distributed between the source of the FRB and the observer, and (ii) the non-zero rest mass of the photon.

\begin{itemize}
    \item Due to the interaction between the plasma and electromagnetic waves, photons with higher frequency $\nu_h$ travel faster than those with lower frequency $\nu_l$, resulting in a dispersion effect. The time delay $\Delta t_{\mathrm {DM}}$ between photons of different frequencies caused by this dispersion effect can be described as \citep{2012hpa..book.....L,BENTUM2017736}
\begin{equation}
\begin{aligned}
\Delta t_{\mathrm {DM}} & =\int \frac{\nu_p^2}{2c}\left ( \nu_l^{-2}-\nu_h^{-2} \right )\mathrm dl \\
& =\frac{e^2}{8\pi^2m_e\epsilon_0c}\left ( \nu_l^{-2}-\nu_h^{-2} \right )\mathrm {DM_{astro}}\;,
\end{aligned}
\end{equation}
where $\nu_p\equiv \sqrt{\frac{n_ee^2}{4\pi^2m_e\epsilon_0}}$ is the plasma frequency with the number density of electrons $n_e$, the charge of an electron $e$, the mass of an electron $m_e$, and the permittivity of vacuum $\epsilon_0$.
Here $\mathrm {DM_{astro}}$ is the dispersion measure (DM) contributed by the plasma, which is defined as the integral of the number density of electrons $n_e$ along the line of propagation, given by $\mathrm {DM_{astro}} \equiv \int n_e \mathrm dl$.

For an extragalactic source, $\mathrm {DM_{astro}}$ can be primarily divided into four components: the contributions from the Milky Way's interstellar medium ($\mathrm {DM_{ ISM}^{MW}}$), the Galactic halo ($\mathrm {DM_{halo}^{MW}}$), the intergalactic medium (IGM; $\mathrm {DM_{IGM}}$), and the host galaxy ($\mathrm {DM_{host}}$). Therefore, $\mathrm {DM_{astro}}$ is given by
\begin{equation}\label{4}
\begin{aligned}
    \mathrm {DM_{astro}}=&\mathrm {DM_{ ISM}^{MW}}+\mathrm {DM_{halo}^{MW}} \\
    &+\mathrm {DM_{IGM}}+\frac{\mathrm {DM_{host,0}}}{1+z}\;,
\end{aligned}
\end{equation}
where the factor $(1+z)$ converts the DM component of the host galaxy in the rest frame, denoted as $\mathrm {DM_{host,0}}$, to the observed value $\mathrm {DM_{host}}$.
\end{itemize}

\begin{itemize}
    \item According to Eq.~(\ref{2}), the time delay $\Delta t_{m_{\gamma}}$ between massive photons of high and low frequencies can be expressed as:
\begin{equation}
\Delta t_{m_{\gamma}}=\frac{1}{2}\left(\frac{m_{\gamma}c^2}{h}\right)^2\left ( \nu_l^{-2}-\nu_h^{-2} \right )H_{\gamma}(z)\;,
\end{equation}
where $H_{\gamma}(z)$ is a newly defined function related to redshift,
\begin{equation}\label{eq:Hr}
    H_{\gamma}(z)=\int_{0}^{z}\frac{\left ( 1+{z}' \right )^{-2}}{H(z')}\mathrm dz'\;,
\end{equation}
where $H(z)$ is the Hubble parameter at redshift $z$. Note that here $H_{\gamma}(z)$ differs from
the dimensionless redshift function that defined in previous works, and it is in units of
$\left [ {\rm km\;s^{-1}\;Mpc^{-1}} \right ]^{-1} $.
\end{itemize}

In our analysis, the total observed time delay is
\begin{equation}
    \Delta t_{\rm obs}=\Delta t_{\mathrm {DM}}+\Delta t_{m_{\gamma}}\;.
\end{equation}
Observationally, the time delay of all FRBs exhibits a frequency dependence of $\nu^{-2}$, while both $\Delta t_{\mathrm {DM}}$ and $\Delta t_{m_{\gamma}}$ follow a $\nu^{-2}$ behavior. Hence, it is natural for us to analogously define the equivalent DM arising from the massive photons as \citep{PhysRevD.95.123010}
\begin{equation}\label{8}
    \mathrm{DM}_{\gamma}(z)\equiv \frac{4\pi^2m_e\epsilon_0c^5}{h^2e^2}H_{\gamma}(z)m_{\gamma}^2\;.
\end{equation}
Thus, the observed DM obtained from fitting the $\nu^{-2}$ behavior in the total frequency-dependent time delay, can be written as:
\begin{equation}
    \mathrm{DM_{obs}}=\mathrm{DM_{astro}}+\mathrm{DM}_{\gamma}\;.
\end{equation}
Once we are able to properly estimate the value of each DM term in Eq.~(\ref{4}), $\mathrm{DM}_{\gamma}$ can then be effectively extracted from $\mathrm{DM_{obs}}$, thereby providing an upper limit on the photon mass $m_{\gamma}$.

The $\mathrm {DM_{ISM}^{MW}}$ term arising from the ionizing medium around our galaxy is well modeled by some galactic electron distribution models. Here we adopt the NE2001 model \citep{2002astro.ph..7156C} for its wide application.
The value of $\mathrm {DM_{halo}^{MW}}$ is challenging to estimate accurately, but it has been expected to lie within the range of $\mathrm{50-80\; pc\;cm^{-3}}$ \citep{2019MNRAS.485..648P,2020MNRAS.496L.106K}. Here, we conservatively adopt a Gaussian prior of $\mathrm{DM_{halo}^{MW}=65\pm 15\; pc\;cm^{-3}}$ \citep{2022MNRAS.515L...1W,2023JCAP...09..025W}.

The DM due to the IGM, $\mathrm {DM_{IGM}}$, depends on the cosmological model and is largely influenced by the number of halos intersected along the propagation path. Due to density perturbations on large-scale structures, it is challenging to calculate the precise value of $\mathrm {DM_{IGM}}$ for individual FRBs. Instead, we typically calculate the average value of $\mathrm {DM_{IGM}}$ using \citep{2014ApJ...783L..35D}
\begin{equation}\label{DMavg}
    \left \langle \mathrm {DM_{IGM}}(z) \right \rangle=\frac{21c\Omega_{b}H_0^2f_{\mathrm {IGM}}}{64\pi Gm_p}H_e(z)\;,
\end{equation}
where $\Omega_b$ is the baryon density parameter at the present day, $H_0$ is the Hubble constant, $f_{\mathrm {IGM}}\simeq 0.83$ is the fraction of baryon in the IGM \citep{1998ApJ...503..518F}, $G$ is the gravitational constant, and $m_p$ is the mass of proton.
The redshift-dependent function $H_e(z)$ (in units of $[{\rm km\;s^{-1}\;Mpc^{-1}}]^{-1}$) is defined by
\begin{equation}\label{eq:He}
    H_e(z)\equiv \int_{0}^{z}\frac{(1+z')}{H(z')}\mathrm dz'\;.
\end{equation}
However, the actual value of $\mathrm {DM_{IGM}}$ may deviate from the average due to the inhomogeneity of IGM. To account for this variability, we employ a one-parameter model to simulate the probability distribution of $\mathrm {DM_{IGM}}$  \citep{2000ApJ...530....1M,2020Natur.581..391M,2021PhLB..82036596W}, which is given by
\begin{equation}\label{pigm}
    P_{\mathrm{IGM}}(\mathrm{DM_{IGM}}|z)=A\Delta ^{-\beta}\exp \left[-\frac{\left ( \Delta ^{-\alpha}-C_0 \right )^2}{2\alpha^2s^2}\right]\;,
\end{equation}
where $\Delta \equiv \mathrm {DM_{IGM}}/\left \langle \mathrm {DM_{IGM}} \right \rangle>0$, $A$ is a normalization constant to ensure the integral of $P_{\mathrm{IGM}}(\mathrm{DM_{IGM}}|z)$ to be unity, $C_0$ is chosen to satisfy $\left \langle \Delta \right \rangle=1$, and $\alpha=\beta=3$ \citep{2020Natur.581..391M}. The fractional standard deviation of $\mathrm {DM_{IGM}}$ is approximated as $s=Fz^{-0.5}$, where $F$ is the free parameter that quantifies the strength of baryon feedback \citep{2020Natur.581..391M}.
Both semi-analytic models and numerical simulations of the IGM and galaxy halos showed that
the probability distribution of $\mathrm {DM_{IGM}}$ can be well fitted by the quasi-Gaussian form (i.e., Eq.~\ref{pigm})
\citep{McQuinn2014, 2020Natur.581..391M}. Therefore, this analytic form is adopted in our analysis.

The $\mathrm{DM_{host}}$ term is contributed by the source environment and the interstellar medium of the host galaxy,  implying that the value of $\mathrm{DM_{host}}$ may vary significantly across different sources. To model this variability, we adopt a log-normal distribution for the probability density function of $\mathrm{DM_{host}}$ with the expression as \citep{2020Natur.581..391M}
\begin{equation}\label{phost}
\begin{aligned}
    P_{\mathrm{host}}\mathrm{(DM_{host}|\mu,\sigma_{host})}&=\frac{1}{\sqrt{2\pi} \mathrm{DM_{host}\sigma_{host}}}\\
    &\times \exp \left[-\frac{\left ( \ln {\mathrm{DM_{host}}}-\mu \right )^2}{2\sigma_{\mathrm{host}}^2}\right]\;,
\end{aligned}
\end{equation}
where $\mu$ and $\sigma_{\mathrm{host}}$ are free parameters which are used to estimate the mean and standard deviation of $\ln {\mathrm{DM_{host}}}$, respectively.

With the analyses stated above, we formulate the joint likelihood function for a sample of localized FRBs as
\begin{equation}\label{eq:likelihood}
    \mathcal{L}=\prod_{i=1}^{N}P_i\left ( \mathrm{DM}_{\mathrm E,i}|z_i \right )\;,
\end{equation}
where $N$ is the number of FRBs and $P_i\left ( \mathrm{DM}_{\mathrm E,i}|z_i \right )$ denotes the likelihood for the $i$-th FRB with the corrected observable DM as
\begin{equation}\label{DME}
\begin{aligned}
    \mathrm{DM_E}&\equiv \mathrm {DM_{obs}}-\mathrm {DM_{halo}^{MW}}-\mathrm {DM_{ ISM}^{MW}}\\
    &=\mathrm {DM_{IGM}}+\mathrm {DM_{host}}+\mathrm {DM_{\gamma}}\;.
\end{aligned}
\end{equation}
For a certain FRB at the redshift of $z_i$, we can estimate the probability distribution of $\mathrm{DM_E}$ with the combination of Eqs.~(\ref{pigm}), (\ref{phost}), and (\ref{DME}), which gives the expression as
\begin{equation}\label{P_DM_E}
\begin{aligned}
    P(&\mathrm{DM_E}|z_i) =\int_0^{\mathrm{DM_E}-\mathrm {DM_{\gamma}}}P_{\rm host}\left(\mathrm {DM_{host}}|\mu,\sigma_{\mathrm{host}}\right)\\
    &\times P_\mathrm{IGM}\left(\mathrm{DM_E}-\mathrm {DM_{host}}-\mathrm {DM_{\gamma}}|z_i\right)\mathrm{dDM_{host}}\;.
\end{aligned}
\end{equation}
Note that by Eq.~(\ref{8}), $\mathrm{DM}_{\gamma}$ is a constant at a given $z_i$.
Denoting $P_X(X)$, $P_Y(Y)$, and $P_Z(Z)$ as the probability density functions of random variables $X$, $Y$, and $Z$, respectively.
The equality $P_Z(Z)=\int_{0}^{Z}P_X(X)P_Y(Z-X)dX$ holds for every non-negative independent random variables $X$, $Y$, and $Z$,
such that $Z=X+Y$.
As $\mathrm{DM}_{\mathrm{IGM}}$ and $\mathrm{DM}_{\mathrm{host}}$ are non-negative independent random variables, by setting $Z=\mathrm{DM}_{\mathrm{E}}-\mathrm{DM}_{\gamma}$, $X=\mathrm{DM}_{\mathrm{host}}$, and $Y=\mathrm{DM}_{\mathrm{IGM}}$, it is easy to derive Eq.~(\ref{P_DM_E}).

By maximizing the likelihood function $\mathcal{L}$, we can place constraints on the parameters in our model, including the rest mass of the photon $m_{\gamma}$. Nevertheless, there are two redshift-dependent functions, $H_{\gamma}(z)$ and $H_e(z)$ (see Eqs.~\ref{eq:Hr} and \ref{eq:He}), to be determined.
In previous works \citep{2016ApJ...822L..15W,2016PhLB..757..548B,2017PhLB..768..326B,2017PhRvD..95l3010S,2019ApJ...882L..13X,2020RAA....20..206W,2021PhLB..82036596W,2023JCAP...01..010C,Lin2023,2023JCAP...09..025W}, the conventional approach is to adopt the flat $\Lambda$CDM cosmological model and replace the denominator
$H(z)$ with $H_0[\Omega_m\left ( 1+z \right )^3+\Omega_{\Lambda}]^{1/2}$. Subsequently, $H_{\gamma}(z)$ and $H_e(z)$ are calculated by treating the cosmological parameters $\Omega_m$ and $\Omega_{\Lambda}$ as fixed values. However, it is crucial to note that the $\Lambda$CDM model is established within the framework of Einstein's theory of GR, which is derived from one of the fundamental postulates--the masslessness of photons. Consequently, adopting this model introduces a logical circularity problem in constraining the rest mass of the photon. To circumvent this issue, it is necessary to employ a cosmology-independent method to describe the relation between the Hubble parameter $H(z)$ and redshift $z$.

\subsection{\label{sec:ANN}Artificial Neural Network}
Gaussian Process (GP) is a widely-used approach for studying cosmology in a model-independent manner. This method assumes that the reconstructed value at a given point follows a Gaussian distribution, and the relationship between function values at different points is characterized by a selected covariance function \citep{2012JCAP...06..036S}. Besides, the reconstructed function of $H(z)$ from GP tends to underestimate errors and is significantly influenced by the prior of the Hubble constant $H_0$ \citep{Wei2017ApJ...838..160W,Wang2017ApJ...847...45W}. In contrast, ANN, inspired by biological neural networks, is a mathematical model primarily used to discover intricate relationships between input and output data. The reconstructed function based on ANN makes no assumptions about the observational data and is entirely data-driven without parameterization of the function. Recently, Ref.~\cite{2020ApJS..246...13W} developed a public code called ReFANN\footnote{\url{https://github.com/Guo-Jian-Wang/refann}} for reconstructing functions from data using ANN. Their reconstructed $H(z)$ function showed no sensitivity to the setting of $H_0$, indicating that reconstructing $H(z)$ from observational data using ANN may be more reliable than using GP. To make a cosmology-independent determination of $H(z)$, in this work we will reconstruct a smooth curve of $H(z)$ from 34 model-independent measurements of $H(z)$ using ReFANN.

The first step in using ReFANN to reconstruct a function from data is to find the best parameter configuration for the model used to train the data. Following the approach of Ref.~\cite{2020ApJS..246...13W}, we determine the optimal network model by minimizing the \textit{risk} \citep{2001astro.ph.12050W}:
\begin{equation}\label{risk}
\begin{aligned}
\mathrm{risk}&=\sum_{i=1}^{N}{\rm Bias}_i^2+\sum_{i=1}^{N}{\rm Variance}_i\\
&=\sum_{i=1}^{N}\left [ H(z_i)-\bar{H}(z_i) \right ]^2+\sum_{i=1}^{N}\sigma^2_{H(z_i)}\;,
\end{aligned}
\end{equation}
where $N$ is the number of sample $H(z)$ and $\bar{H}(z_i)$ is the fiducial value of $H(z)$ at redshift $z_i$. For our purposes, we only consider the number of hidden layers and neurons in the hidden layer. Through experimentation, we determined that the optimal model configuration involves a single hidden layer with 4096 neurons. Additionally, to reduce sensitivity to initialization and stabilize the distribution among variables, we employ batch normalization \citep{2015arXiv150203167I} in our network model, with the batch size set to half of the number of $H(z)$ data points. We also use the Exponential Linear Unit \citep{2015arXiv151107289C} as the activation function, with the hyperparameter $\alpha$ set to be 1. Furthermore, we utilize the Adam optimizer \citep{2014arXiv1412.6980K} to update the network parameters in each iteration of the training process.

Subsequently, we train this optimal network model with the observational $H(z)$ data. This training process provides us with a predicted value of $H(z_i)$ and its associated error $\sigma_{H(z_i)}$ at a given redshift $z_i$. This predicted function represents an approximate reconstruction of $H(z)$ based on the ANN model trained with observational data.

\begin{table}[htb!]
\centering
\caption{34 $H(z)$ measurements obtained with the CC method. \label{CCH}}
\setlength{\tabcolsep}{5pt}{
\begin{tabular}{cccc}
\hline\hline
$z$ & $H(z)$& $\sigma$ & \textrm{Refs.}\\
    & ($\rm km\;s^{-1}\;Mpc^{-1}$) & ($\rm km\;s^{-1}\;Mpc^{-1}$) &  \\
\hline
0.09 & 69 & 12 & \cite{2003ApJ...593..622J}\\
\hline
0.17 & 83 & 8 & \multirow{8}*{\cite{2005PhRvD..71l3001S}}\\
0.27 & 77 & 14 &  \\
0.4 & 95 & 17 &  \\
0.9 & 117 & 23 &  \\
1.3 & 168 & 17 &  \\
1.43 & 177 & 18 &  \\
1.53 & 140 & 14 &  \\
1.75 & 202 & 40 &  \\
\hline
0.48 & 97 & 62 & \multirow{2}*{\cite{2010JCAP...02..008S}}\\
0.88 & 90 & 40 &  \\
\hline
0.1791 & 75 & 4 & \multirow{8}*{\cite{2012JCAP...08..006M}}\\
0.1993 & 75 & 5 &  \\
0.3519 & 83 & 14 &  \\
0.5929 & 104 & 13 &  \\
0.6797 & 92 & 8 &  \\
0.7812 & 105 & 12 &  \\
0.8754 & 125 & 17 &  \\
1.037 & 154 & 20 &  \\
\hline
0.07 & 69 & 19.6 & \multirow{4}*{\cite{2014RAA....14.1221Z}}\\
0.12 & 68.2 & 26.2 &  \\
0.2 & 72.9 & 29.6 &  \\
0.28 & 88.8 & 36.6 &  \\
\hline
1.363 & 160 & 33.6 & \multirow{2}*{\cite{2015MNRAS.450L..16M}}\\
1.965 & 186.5 & 50.4 &  \\
\hline
0.3802 & 83 & 13.5 & \multirow{5}*{\cite{2016JCAP...05..014M}}\\
0.4004 & 77 & 10.2 &  \\
0.4247 & 87.1 & 11.2 &  \\
0.4497 & 92.8 & 12.9 &  \\
0.4783 & 80.9 & 9 &  \\
\hline
0.47 & 89 & 49.6 & \cite{2017MNRAS.467.3239R}\\
\hline
0.75 & 98.8 & 33.6 & \cite{2022ApJ...928L...4B}\\
\hline
0.8 & 113.1 & 25.22 & \cite{2023ApJS..265...48J}\\
\hline
1.26 & 135 & 65 & \cite{2023AA...679A..96T}\\
\hline\hline
\end{tabular}}
\end{table}

\subsection{\label{sec:H(z)}Hubble parameter $H(z)$ Data}
The Hubble parameter $H(z)$, which quantifies the expansion rate of the universe as a function of time (or equivalently redshift), has been extensively used for exploring the nature of dark energy and testing modified theories of gravity.
$H(z)$ can be measured using two main methods. One approach is based on the detection of the radial baryon acoustic oscillation features \citep{2009MNRAS.399.1663G,2012MNRAS.425..405B,2013MNRAS.429.1514S}. However, this method relies on assumptions about the underlying cosmological model, typically the $\Lambda$CDM model. Therefore, in our analysis, we focus exclusively on $H(z)$ measurements obtained using another method, namely the cosmic chronometers (CC) method \citep{2002ApJ...573...37J}. This method is model-independent as it relies on minimal assumptions, primarily the use of a Friedmann-Lema\^{\i}tre-Robertson-Walker metric. In this method, by combining the definition of $H(z)\equiv \dot{a}/a$ and the relation between the scale factor $a(t)$ and redshift, $1+z=1/a(t)$, $H(z)$ can be rewritten as
\begin{equation}
    H(z)=-\frac{1}{1+z}\frac{\mathrm dz}{\mathrm dt}\;.
\end{equation}
This means that one can achieve a model-independent determination of $H(z)$ by calculating the differential age of the universe using passively evolving galaxies at various redshifts. We compile the latest 34 CC $H(z)$ measurements in Table~\ref{CCH}, covering the redshift range of $0.07<z<1.965$.

\begin{figure}[t!]
\vskip-0.2in
\centering
\includegraphics[width=1.0\linewidth]{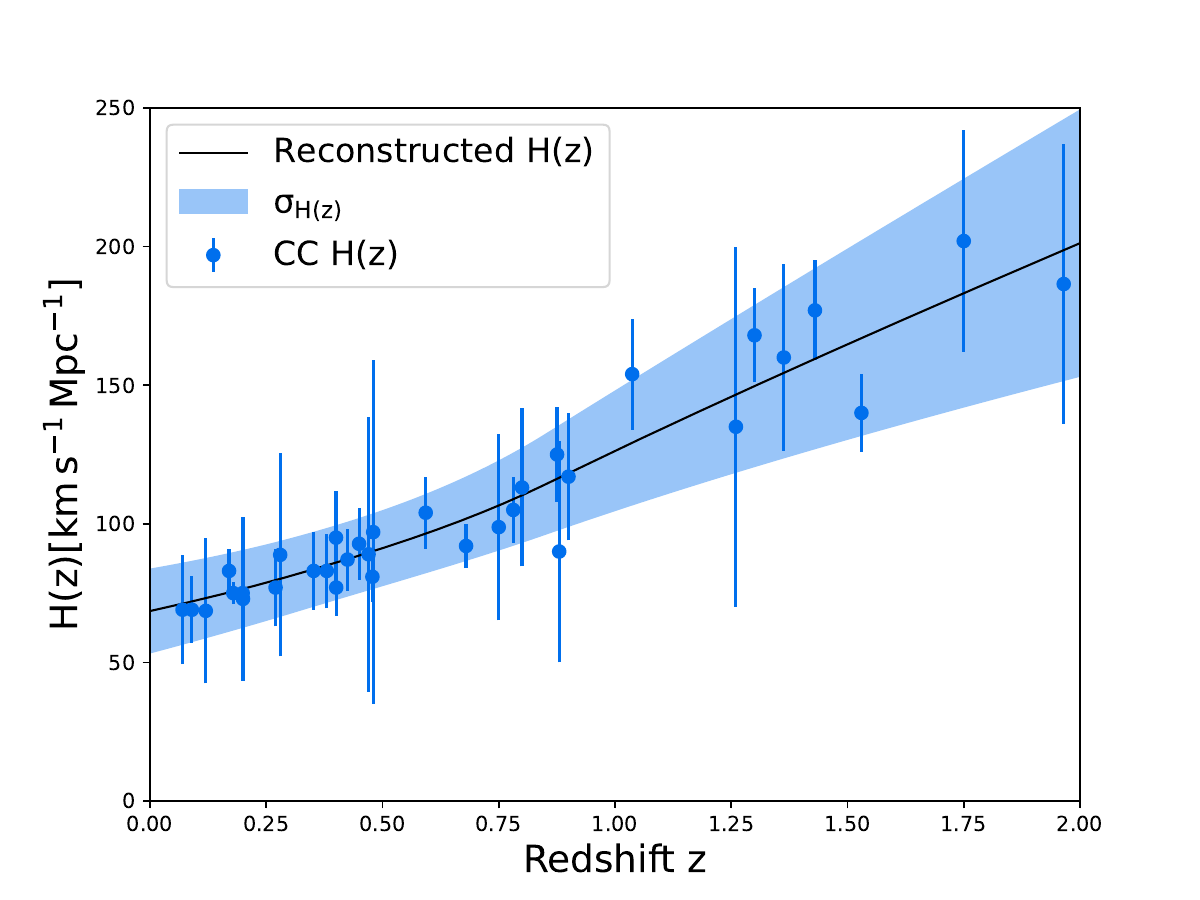}
\caption{\label{rec_H}Reconstruction of the Hubble parameter $H(z)$ from 34 CC $H(z)$ measurements using ANN. The shaded area corresponds to the 1$\sigma$ confidence region of the reconstruction. The blue dots with error bars depict the observational $H(z)$ data.}
\end{figure}

\begin{figure}[htb!]
\vskip-0.2in
\centering
\includegraphics[width=0.98\linewidth]{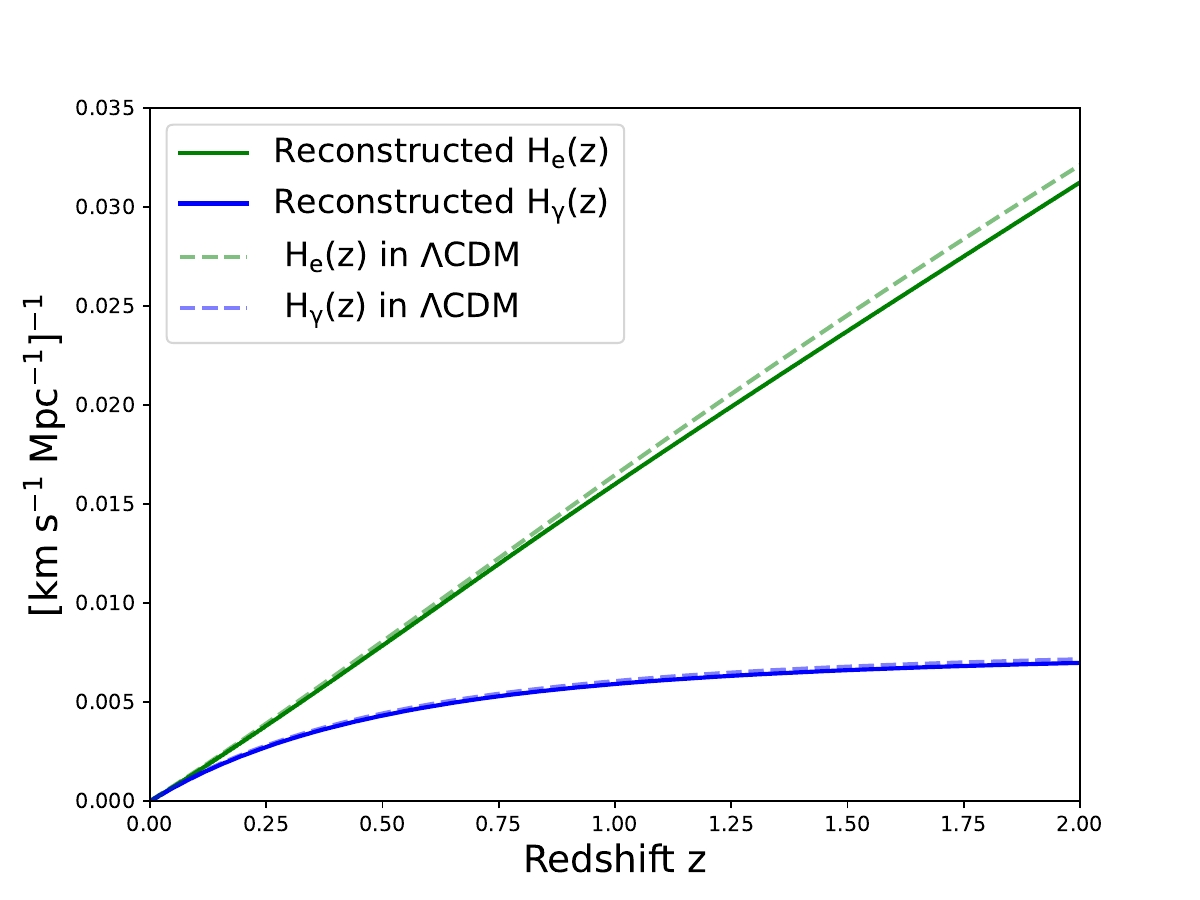}
\caption{\label{He_Hgamma} The redshift dependence of the reconstructed $H_{\gamma}(z)$ and $H_e(z)$ functions
(solid lines). The corresponding theoretical curves (dashed lines) derived from the flat $\Lambda$CDM model with
$\Omega_m=0.315$ and $H_0=67.4\;\rm km\;s^{-1}\;Mpc^{-1}$ are also shown.}
\end{figure}

Having obtained the dataset of $H(z)$, we adopt ANN to reconstruct the $H(z)$ function within the redshift range of $0<z<2$,
and the results are shown in Fig.~\ref{rec_H}. The black line represents the best fit, and the blue shaded area indicates
the 1$\sigma$ confidence region of the reconstructed function.
The redshift-dependent $H_{\gamma}(z)$ and $H_e(z)$ functions can then be derived by integrating the reconstructed $H(z)$
function with respect to redshift. As shown in Fig.~\ref{He_Hgamma}, the reconstructed $H_{\gamma}(z)$ and $H_e(z)$ functions exhibit significant differences in behavior as a function of redshift $z$. Such distinctions are crucial for breaking parameter degeneracy and enhancing sensitivity in testing the photon mass when a few redshift measurements of
FRBs are available \citep{2016PhLB..757..548B,2017PhLB..768..326B,2017PhRvD..95l3010S,2020RAA....20..206W,2017AdSpR..59..736B}.
For comparison, we also plot $H_{\gamma}(z)$ and $H_e(z)$ functions within the framework of the flat $\Lambda$CDM
model using Planck 2018 parameters ($\Omega_m=0.315$ and $H_0=67.4\;\rm km\;s^{-1}\;Mpc^{-1}$) \citep{2020AA...641A...6P}.
The corresponding theoretical curves are presented in Fig.~\ref{He_Hgamma} as dashed lines.
It is evident that the reconstructions of $H_{\gamma}(z)$ and $H_e(z)$ are roughly consistent with the predictions of
the standard $\Lambda$CDM model, implying that the ANN method can offer a reliable reconstructed function from the
observational data.

\subsection{\label{sec:frb}Fast Radio Burst Data}
Expanding upon the sample of 23 localized FRBs utilized in Ref.~\cite{2023JCAP...09..025W}, we incorporate 9 new localized FRBs recently detected by the 110-antenna Deep Synoptic Array \cite{2023arXiv230703344L}. Note that another two new FRBs with redshift measurements, FRB 20220319D and FRB 20220914A, are not included in our sample.
FRB 20220319D exhibits an observed $\mathrm{DM_{obs}}$ value of $110.98\;\mathrm{pc\;cm^{-3}}$, while the $\mathrm{DM_{ISM}^{MW}}$ value calculated using the NE2001 model stands at $133.3\;\mathrm{pc\;cm^{-3}}$, leading to its exclusion from our analysis. Furthermore, FRB 20220914A is omitted due to its notable deviation from the expected $\mathrm{DM_{IGM}}$--$z$ relation outlined in Eq.~(\ref{DMavg}). The total 32 localized FRBs, including their respective redshifts, $\mathrm{DM_{obs}}$, and $\mathrm{DM_{ISM}^{MW}}$, are provided in Table~\ref{FRBs}.

\begin{table}[htb!]
\centering
\caption{Properties of 32 localized FRBs. \label{FRBs}}
\setlength{\tabcolsep}{5pt}{
\begin{tabular}{lcccc}
\hline\hline
\textrm{Name} & $z$ & $\mathrm{DM_{obs}}$ & $\mathrm{DM_{ISM}^{MW}}$ & \textrm{Refs.}\\
              &     & $(\mathrm{pc \, cm^{-3}})$ & $(\mathrm{pc \, cm^{-3}})$ & \\
\hline
FRB 20121102 & 0.19273 & 557  & 188.0 & \cite{2017Natur.541...58C}\\
FRB 20180301 & 0.3304 & 536  & 152.0 &  \cite{2022AJ....163...69B}\\
FRB 20180916 & 0.0337 & 348.76  & 200.0 & \cite{2020Natur.577..190M}\\
FRB 20180924 & 0.3214 & 361.42  & 40.5 & \cite{2019Sci...365..565B}\\
FRB 20181112 & 0.4755 & 589.27  & 102.0 & \cite{2019Sci...366..231P}\\
FRB 20190102 & 0.291 & 363.6  & 57.3 & \cite{Bhandari2020}\\
FRB 20190523 & 0.66 & 760.8  & 37.0 & \cite{Ravi2019} \\
FRB 20190608 &  0.1178 & 338.7  & 37.2 & \cite{Chittidi2021} \\
FRB 20190611 & 0.378 & 321.4  & 57.83 & \cite{Heintz2020} \\
FRB 20190614 & 0.6 & 959.2 & 83.5 & \cite{Law2020} \\
FRB 20190711 & 0.522 & 593.1 & 56.4 & \cite{Heintz2020} \\
FRB 20190714 & 0.2365 & 504 & 38.0 & \cite{Heintz2020} \\
FRB 20191001 & 0.234 & 506.92 & 44.7 & \cite{Heintz2020} \\
FRB 20191228 & 0.2432 & 297.5 & 33.0 & \cite{2022AJ....163...69B} \\
FRB 20200430 & 0.16 & 380.1 & 27.0 & \cite{Heintz2020} \\
FRB 20200906 & 0.3688 & 577.8 & 36.0 & \cite{2022AJ....163...69B} \\
FRB 20201124 & 0.098 & 413.52 & 123.2 & \cite{Ravi2022} \\
FRB 20210117 & 0.2145 & 730 & 34.4 & \cite{James2022} \\
FRB 20210320 & 0.27970 & 384.8 & 42 & \cite{James2022} \\
FRB 20210807 & 0.12927 & 251.9 & 121.2 & \cite{James2022}\\
FRB 20211127 & 0.0469 &234.83 & 42.5 & \cite{James2022} \\
FRB 20211212 & 0.0715 & 206 & 27.1 & \cite{James2022} \\
FRB 20220207C & 0.043040 & 262.38 & 79.3 & \cite{2023arXiv230703344L}\\
FRB 20220307B & 0.28123 & 499.27 & 135.7 & \cite{2023arXiv230703344L}\\
FRB 20220310F & 0.477958 & 462.24 & 45.4 & \cite{2023arXiv230703344L}\\
FRB 20220418A & 0.622000 & 623.25 & 37.6 & \cite{2023arXiv230703344L}\\
FRB 20220506D & 0.30039 & 396.97 & 89.1 & \cite{2023arXiv230703344L}\\
FRB 20220509G & 0.089400 & 269.53 & 55.2 & \cite{2023arXiv230703344L}\\
FRB 20220610A & 1.016 & 1457.624 & 31 & \cite{Ryder2022} \\
FRB 20220825A & 0.241397 & 651.24 & 79.7 & \cite{2023arXiv230703344L}\\
FRB 20220920A & 0.158239 & 314.99 & 40.3 & \cite{2023arXiv230703344L}\\
FRB 20221012A & 0.284669 & 441.08 & 54.4 & \cite{2023arXiv230703344L}\\
\hline\hline
\end{tabular}}
\end{table}

\section{\label{sec:results}Results}
We explore the posterior probability distributions of the free parameters by maximizing the joint likelihood function $\mathcal{L}$ (Eq.~\ref{eq:likelihood}) using emcee, an affine-invariant Markov Chain Monte Carlo (MCMC) ensemble sampler implemented in Python \citep{2013PASP..125..306F}.
The free parameters of our model now include the baryon density parameter $\Omega_b h_{0}^2$ (where $h_{0} \equiv H_0/{\rm (100\,km\,s^{-1}\,Mpc^{-1})}$), the DM contribution from the Milky Way's halo $\mathrm {DM_{halo}^{MW}}$, the parameters related to the probability distributions for $\mathrm {DM_{IGM}}$
and $\mathrm{DM_{host}}$ (i.e., $F$, $\mu$, and $\sigma_{\mathrm{host}}$), and the photon mass $m_{\gamma}$.
In our MCMC analysis, we set uniform priors on $\Omega_{b}h_{0}^2\in[0.01,\;1]$, $F\in[0.01,\;0.5]$, $e^{\mu}\in[20,\;200]$ $\mathrm{pc\;cm^{-3}}$, $\sigma_{\mathrm{host}}\in[0.2,\;2]$, and $m_{\gamma}\in[0,\;10^{-49}]$ $\mathrm{kg}$.
For $\mathrm{DM_{halo}^{MW}}$, we set a Gaussian prior, $\mathrm{DM_{halo}^{MW}=65\pm 15\; pc\;cm^{-3}}$, within the wide $3\sigma$ range of $[20,\;110]$ $\mathrm{pc\;cm^{-3}}$.
To incorporate the error of the ANN reconstruction into our analysis, at each MCMC step,
we sample the Hubble rate $\tilde{H}(z)$ function according to the Gaussian distribution
$\tilde{H}(z)=\mathcal{N}(H(z),\,\sigma_{H(z)})$ with mean $H(z)$ and standard deviation $\sigma_{H(z)}$.
Here $H(z)$ is the best-fit function reconstructed by ANN and $\sigma_{H(z)}$ is the $1\sigma$ error of the reconstructed $H(z)$ function.

\begin{figure*}[ht!]
\centering
\includegraphics[width=0.8\linewidth]{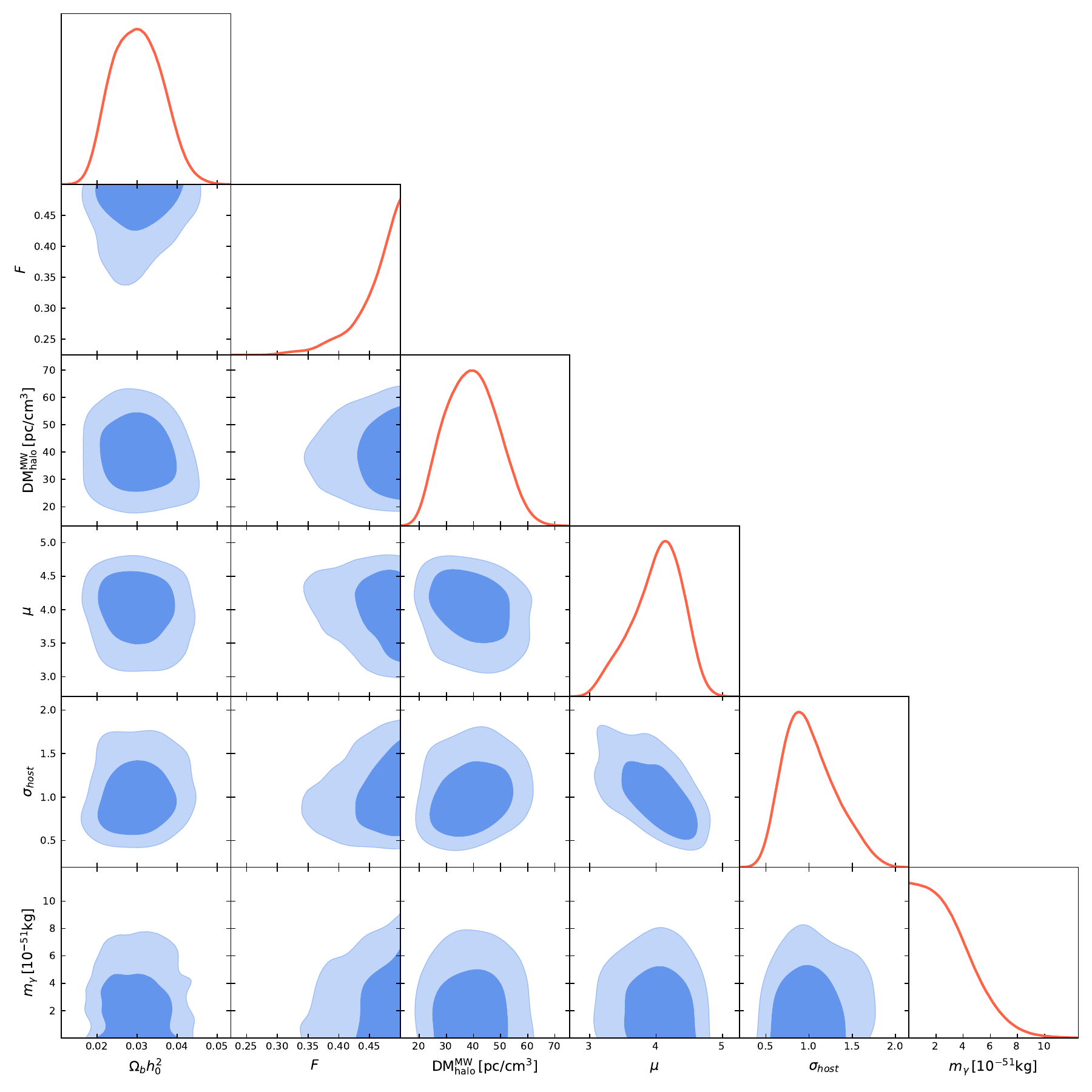}
\caption{Plots of 1D marginalized probability distributions and 2D $1-2\sigma$ confidence contours for the parameters $\Omega_b h_{0}^2$, $F$, $\mathrm {DM_{halo}^{MW}}$, $\mu$, $\sigma_{\mathrm{host}}$, and $m_{\gamma}$, constrained by 32 localized FRBs.
\label{fig:results}}
\end{figure*}

The $1\sigma$ constraint results for these six parameters are summarized in Table~\ref{tab:results}. For the photon mass $m_{\gamma}$, both the $1\sigma$ and $2\sigma$ upper limits are presented. Posterior distributions and $1-2\sigma$ confidence regions for these parameters are displayed in Fig.~\ref{fig:results}. Notably, the constraints on $m_{\gamma}$ are determined as
\begin{equation}
    m_{\gamma} \le  3.5 \times 10^{-51} \, \rm{kg} \simeq 2.0 \times 10^{-15} \, \rm{eV/c^2}
\end{equation}
and
\begin{equation}
    m_{\gamma} \le  6.5 \times 10^{-51} \, \rm{kg} \simeq 3.6 \times 10^{-15} \, \rm{eV/c^2}
\end{equation}
at the $1\sigma$ and $2\sigma$ confidence level, respectively. Meanwhile, we find that the baryon density
parameter is optimized to be $\Omega_b h_{0}^2=0.030_{-0.007}^{+0.006}$, which is compatible with the value inferred from
Planck 2018 ($\Omega_b h_0^2 = 0.0224 \pm 0.0001$) at the $1.2\sigma$ confidence level \citep{2020AA...641A...6P}.

To investigate the impact of the prior assumption of $\mathrm {DM_{halo}^{MW}}$ on our results,
we also perform a parallel comparative analysis of the FRB data using a flat prior on
$\mathrm {DM_{halo}^{MW}}\in[20,\;110]$ $\mathrm{pc\;cm^{-3}}$.
The corresponding resulting constraints on all parameters are also reported in Table~\ref{tab:results}.
Comparing these inferred parameters with those obtained from the Gaussian prior (see line 1 in
Table~\ref{tab:results}), it is clear that, except for the best-fit value of $\mathrm {DM_{halo}^{MW}}$,
the prior assumption of $\mathrm {DM_{halo}^{MW}}$ only has a minimal influence on our results.

\begin{table*}[ht!]
\centering
\caption{The $1\sigma$ constraints for all parameters with different priors on $\mathrm {DM_{halo}^{MW}}$. Both the $1\sigma$ and $2\sigma$ upper limits for the photon mass $m_{\gamma}$ are provided.\label{tab:results}}
\setlength{\tabcolsep}{12pt}{
\begin{tabular}{ccccccc}
\hline\hline
Priors of $\mathrm {DM_{halo}^{MW}}$ & $\Omega_b h_{0}^2$ & $F$ & $\mathrm {DM_{halo}^{MW}}$ & $\mu$ & $\sigma_{\mathrm{host}}$ & $m_{\gamma}$\\
  &   &  & $(\mathrm{pc\;cm^{-3}})$ &   &   & $(10^{-51}\,{\rm kg})$\\
\hline
Gaussian prior & $0.030_{-0.007}^{+0.006}$ & $\geq 0.449$ & $39_{-10}^{+10}$ & $4.02_{-0.29}^{+0.46}$ & $1.01_{-0.37}^{+0.22}$ & $\leq 3.5\; (\leq 6.5)$  \\
Flat prior & $0.031_{-0.006}^{+0.006}$ & $\geq 0.451$ & $\leq 32$ & $4.07_{-0.30}^{+0.48}$ & $0.99_{-0.37}^{+0.21}$ & $\leq 3.8\; (\leq 7.2)$  \\
\hline\hline
\end{tabular}}
\end{table*}

\section{\label{sec:discussions}Discussion and conclusions}
As a matter of fact, it is impossible to prove experimentally that the rest mass of a photon is strictly zero.
According to the uncertainty principle of quantum mechanics, the ultimately measurable order of magnitude for
the photon mass is $m_{\gamma}\approx \hbar/(\Delta tc^2)\approx 10^{-69}\,\rm {kg}$, where $\hbar$ is the reduced Planck constant and $\Delta t\approx 10^{10}\,\rm {years}$ is the age of our universe. However, due to the immense importance of Maxwell's theory and Einstein's theory of relativity, it is still intellectually stimulating and scientifically significant to approach this ultimate limit by various experiments.

FRBs provide the current best celestial laboratory to test the photon mass $m_{\gamma}$ via the dispersion method.
Constraining $m_{\gamma}$ with cosmological FRBs, however, one has to know the cosmic expansion rate $H(z)$.
In all previous studies, the required $H(z)$ information is estimated within the standard $\Lambda$CDM cosmological model. Such $H(z)$ estimations would involve a circularity problem in constraining $m_{\gamma}$,
since $\Lambda$CDM itself is built on the framework of GR and GR embraces the postulate of the constancy of light speed.
In this work, aiming to overcome the circularity problem, we have employed an ANN technology to reconstruct a
cosmology-independent $H(z)$ function from the discrete CC $H(z)$ data.

By combining the DM--$z$ measurements of 32 localized FRBs with the reconstructed $H(z)$ function from 34 CC $H(z)$ data,
we have placed the first cosmology-independent photon mass limit. Our results
show that the $1\sigma$ and $2\sigma$ confidence-level upper limits on the photon mass are
$m_{\gamma} \le  3.5 \times 10^{-51} \, \rm{kg}$ (or equivalently $m_{\gamma} \le 2.0 \times 10^{-15} \, \rm{eV/c^2}$)
and $m_{\gamma} \le  6.5 \times 10^{-51} \, \rm{kg}$ (or equivalently $m_{\gamma} \le  3.6 \times 10^{-15} \, \rm{eV/c^2}$), respectively. Previously, under the assumption of fiducial $\Lambda$CDM cosmology, Ref.~\citep{2021PhLB..82036596W} obtained an upper limit of $m_{\gamma} \le  3.1 \times 10^{-51} \, \rm{kg}$ at the $1\sigma$ confidence level by using a catalog of 129 FRBs (most of them without redshift measurement, and the observed $\mathrm{DM_{obs}}$ values were used to estimate the pseudo redshifts). Ref.~\citep{Lin2023} obtained $m_{\gamma} \le  4.8 \times 10^{-51} \, \rm{kg}$ ($1\sigma$) by analyzing a sample of 17 localized FRBs in the flat $\Lambda$CDM model. Ref.~\citep{2023JCAP...09..025W} obtained $m_{\gamma} \le  3.8 \times 10^{-51} \, \rm{kg}$ ($1\sigma$) for flat $\Lambda$CDM using a sample of 23 localized FRBs.
Despite not assuming a specific cosmological model, the precision of our constraint from 32 localized FRBs is
comparable to these previous results. Most importantly, this highlights the validity of our approach and suggests that as the number of CC $H(z)$ measurements increases, we can expect even more reliable model-independent tests of the photon mass.

\begin{acknowledgments}
We are grateful to the anonymous referees for their helpful comments.
This work is partially supported by the National SKA Program of China (2022SKA0130100),
the National Natural Science Foundation of China (grant Nos. 12373053, 12321003, and
12041306), the Key Research Program of Frontier Sciences (grant No. ZDBS-LY-7014)
of Chinese Academy of Sciences, International Partnership Program of Chinese Academy of Sciences
for Grand Challenges (114332KYSB20210018), the CAS Project for Young Scientists in Basic Research
(grant No. YSBR-063), the CAS Organizational Scientific Research Platform for National Major
Scientific and Technological Infrastructure: Cosmic Transients with FAST, and the Natural Science
Foundation of Jiangsu Province (grant No. BK20221562).
\end{acknowledgments}

\nocite{*}


%

\end{document}